\documentclass[prl,floatfix,twocolumn,superscriptaddress]{revtex4-1}

\usepackage{graphicx}
\usepackage{subfigure}
\usepackage{stdclsdv}
\usepackage{array}
\usepackage{amsmath}
\usepackage{amsfonts}

\usepackage{hyperref}
\usepackage{xcolor}
\hypersetup{
    colorlinks,
    linkcolor={red!50!black},
    citecolor={blue!50!black},
    urlcolor={blue!80!black}
}

\newcommand{\av}[1]{\langle  #1 \rangle}
\newcommand{\Fig}[1]{Fig.~\ref{#1}}

\newcommand{\Eq}[1]{Eq.~\ref{#1}}

\begin{document}

\title{ Stoichiometry controls the dynamics of  liquid condensates of associative proteins }

\author{Pierre Ronceray}
\thanks{Equal contribution.}
\affiliation{Center for the Physics of Biological Function, Princeton University}

\author{Yaojun Zhang}
\thanks{Equal contribution.}
\affiliation{Center for the Physics of Biological Function, Princeton University}

\author{Xichong Liu}
\affiliation{Department of Chemical and Biological Engineering, Princeton University}
\affiliation{Stanford University School of Medicine}

\author{Ned S. Wingreen}
\email{wingreen@princeton.edu}
\affiliation{Department of Molecular Biology, Princeton University}
\affiliation{Lewis-Sigler Institute for Integrative Genomics, Princeton University}

\begin{abstract}
  Multivalent associative proteins with strong complementary
  interactions play a crucial role in phase separation of
  intracellular liquid condensates. We study the internal dynamics of
  such “bond-network” condensates comprised of two complementary
  proteins via scaling analysis and molecular dynamics. We find that
  when stoichiometry is balanced, relaxation slows down dramatically
  due to a scarcity of alternative partners following a bond
  break. This microscopic slow-down strongly affects the bulk
  diffusivity, viscosity and mixing, which provides a means to
  experimentally test our predictions.
\end{abstract}

\maketitle

Protein-rich liquid condensates, also known as membraneless
organelles, have recently emerged as an important paradigm for
intracellular
organization~\cite{brangwynne_germline_2009,brangwynne_phase_2013,banani_biomolecular_2017}.
Several distinct molecular mechanisms involved in condensate phase
separation have been characterized~\cite{dignon_biomolecular_2020},
including weak interactions between intrinsically disordered regions
of proteins, interactions with RNA and DNA, and specific
protein-to-protein complementary interactions. Here we focus on the
latter mechanism, often described in terms of ``sticker-and-spacer''
models~\cite{choi_physical_2020}, where strongly interacting
complementary ``stickers'' are separated by flexible ``spacers'', which have
little to no interactions. In a simple case, only two species are
involved with complementary sticker domains
(\Fig{fig:schematic}a), and the phase-separated liquid consists of a
dynamically rearranging network of these bound domains
(\Fig{fig:schematic}b). This paradigm of a binary mixture of
complementary proteins has been observed in membraneless organelles
such as the algal pyrenoid~\cite{freeman_rosenzweig_eukaryotic_2017},
as well as in artificial protein condensates such as SUMO-SIM
assemblies~\cite{banani_compositional_2016}.

Recent studies show that such binary liquids characterized by strong
complementary interactions differ in their properties from usual,
non-biological liquids: for instance, their valence sensitively
controls their phase boundary through a ``magic number''
effect~\cite{freeman_rosenzweig_eukaryotic_2017,xu_rigidity_2020,zhang_decoding_2020},
and they can exhibit long-lived metastable clusters prior to
macroscopic phase separation following a
quench~\cite{ranganathan_dynamic_2020}.  Little is known however about
the bulk dynamical properties of these liquids. It is expected that
these liquids will inherit some properties of associative polymers---a
class of materials characterized by long chains with sparse sticky
sites~\cite{rubinstein_solutions_1997}.  In these materials,
relaxation is slowed down by the attachment-detachment dynamics of
binding sites, resulting in \emph{sticky
  reptation}~\cite{zhang_dynamics_2018}. However, the corresponding
role of attachment-detachment dynamics has not yet been considered in
liquid protein condensates.

In this Letter, we study the bulk relaxation mechanisms of liquids
consisting of a binary mixture of multivalent complementary proteins
(\Fig{fig:schematic}a-b). With theory and simulations, we show that
even in such simple systems, the strong specificity of interactions
results in a finely tuned response to changes in composition---a
property that cells might exploit to dynamically adapt the mixing
properties of condensates.  We first present a simple kinetic model
that predicts a strong dependence of the local relaxation time of
bonds on the composition of the liquid: at equal stoichiometry of
complementary domains, we anticipate a sharp peak in the relaxation
time. We then employ molecular dynamics simulations to confirm these
predictions and show their striking consequences for the bulk diffusivity and
the overall viscosity of the liquid. Finally, we demonstrate that this
effect quantitatively and qualitatively affects the mixing dynamics of
droplets of different compositions, and propose experimental ways to
test our theoretical predictions.

\subsection*{Kinetic model for the bond relaxation.}

\begin{figure}[b]
  \centering
  \includegraphics[width=\columnwidth]{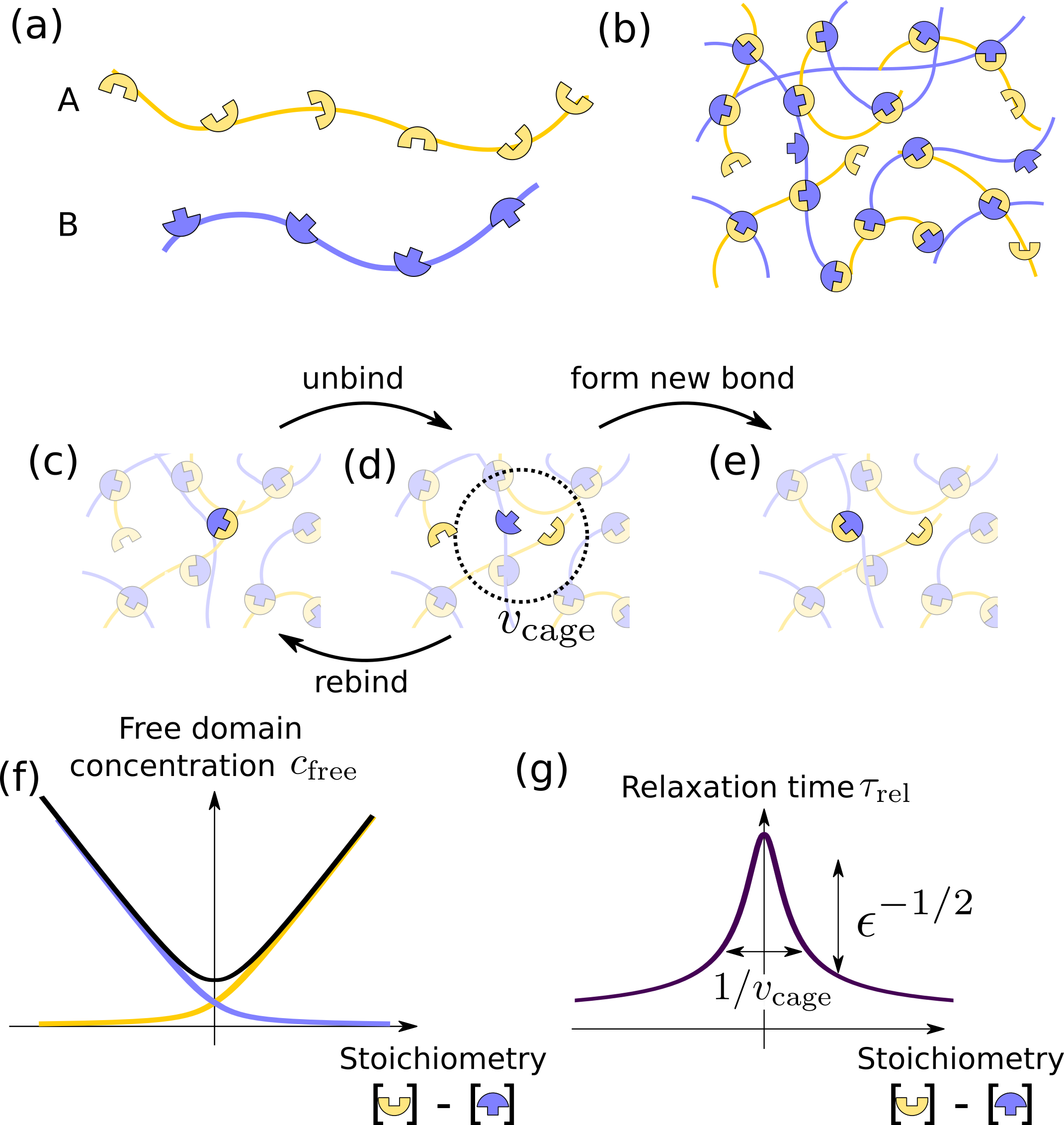}
  \caption{ \textbf{Stoichiometry controls the bond relaxation time of
      multivalent associative proteins.}  \textbf{(a)} Sketch of
    associative multivalent proteins, with complementary domains
    separated by flexible linkers. \textbf{(b)} Strong yet
    reversible binding between proteins leads them to condense into a
    network with most bonds satisfied. \textbf{(c-e)} Schematic of the
    bond relaxation mechanism. When two initially bound domains (c)
    unbind, the two are caged in a small volume $v_\mathrm{cage}$
    (d). Two events can then occur: the initially bound domains can
    rebind, or, if a free domain is within reach, a new bond may form
    (e) which is the system's basic relaxation mechanism.
    \textbf{(f)} Fraction of unbound domains (\Eq{eq:cfree}) of both
    types as a function of stoichiometry difference. \textbf{(g)}
    Relaxation time (\Eq{eq:tau}) corresponding to the process of
    unbinding and then rebinding with a new partner (c-e), as a
    function of stoichiometry difference. Here
    $\epsilon = e^{-\Delta F}$.}
  \label{fig:schematic}
\end{figure}

We consider the dense phase of multivalent proteins of two different
types, denoted A and B (\Fig{fig:schematic}a), where each domain can
bind to one and only one domain of the complementary type.  The free
energy favoring formation of such a bond is $\Delta F$, with a
corresponding unbinding Arrhenius factor $\epsilon = \exp(-\Delta F)$
(we set the thermal energy $k_B T=1$ throughout). We consider the
strong-binding regime, \emph{i.e.}  $\epsilon \ll 1$. In this regime,
the system at any time looks like a gel-forming network with most
bonds between domains satisfied (\Fig{fig:schematic}b). However, over
sufficiently long times, bonds still break and rearrange, the system
relaxes, and the system can flow as a liquid. We investigate here the
dependence of this relaxation time on the Arrhenius factor $\epsilon$
and on the composition of the liquid.

In the strong-binding regime, local relaxation is controlled by
individual bond breaking (\Fig{fig:schematic}c). This process is
slow and thermally activated,  occurring at a dissociation rate
$k_d = \epsilon / \tau_0 $ where $\tau_0$ is a microscopic relaxation
time, and these events are rapidly followed by  rebinding. However, the
two newly unbound complementary domains are part of the network, and
thus are not free: they remain confined and diffuse only in
a small volume $v_\mathrm{cage}$ around their initial position
(\Fig{fig:schematic}d). This caging volume is determined by the length
and flexibility of linkers. Subsequent to a bond breaking, there is
therefore a high probability the two former partners will rebind to
each other, thus negating the effect of the bond break on
system relaxation. Only if either of the two finds a new, unbound
complementary domain within the cage volume (\Fig{fig:schematic}e)
does the initial break contribute to system relaxation and liquidity.

If we denote by $p$ the probability that either domain finds a new
partner, the effective relaxation time can thus be approximated as
$\tau_\mathrm{rel} = 1/ (p k_d)$.  To estimate the probability $p$, we
note that if there are on average $n$ free domains in the volume
$v_\mathrm{cage}$, the probability of finding a new partner prior to
rebinding to the former can be approximated as $p = n/(1+n)$. We can
then express $n= v_\mathrm{cage}c_\mathrm{free}$ in terms of the
concentration $c_\mathrm{free}= c_{\mathrm{A}} + c_{\mathrm{B}}$ of
unbound domains in the system, where we denote by $c_\mathrm{A}$ and
$c_\mathrm{B}$ the respective concentration of free domains of each
type. We define the stoichiometry difference $\delta = c_\mathrm{A} - c_\mathrm{B}$ as the
difference between these concentrations (which depends only on the
overall composition, not on the fraction bound), and $c_{\mathrm{AB}}$
as the concentration of bound domain pairs. We assume that the linkers
are sufficiently flexible to consider the binding state of each domain
of a protein as independent of the others, and thus treat the
binding-unbinding process as a well-mixed solution. The dissociation
equilibrium reads
$K_d = c_\mathrm{A} c_\mathrm{B} / c_{\mathrm{AB}}$, with $K_d$ the
dissociation constant. We thus have:
\begin{equation}
  \label{eq:cfree}
  c_\mathrm{free} = \sqrt{\delta^2 + 4 K_d c_{\mathrm{AB}}}.
\end{equation}
The concentration of free monomers thus exhibits a global minimum at
$\delta = 0$ (\Fig{fig:schematic}f).

We relate the dissociation constant to the Arrhenius factor for
unbinding, writing $K_d = \epsilon / v_0$ where $v_0$ is a molecular
volume. Indeed, $K_d = k_d / k_a$ where the dissociation rate
$k_d=\epsilon /\tau_0$ is proportional to the Arrhenius factor, assuming that the association rate $k_a$  is independent of the
binding strength. We can thus express the relaxation time as:
\begin{equation}
  \label{eq:tau}
  \tau_\mathrm{rel} = \frac{\tau_0}{\epsilon} \left( 1+ \frac{1}{ v_\mathrm{cage} \sqrt{\delta^2 + 4 \epsilon c_{AB}/v_0 } } \right).
\end{equation}
When $n\ll 1$, \emph{i.e.} when there are few available partners
within reach of a domain, the second term in \Eq{eq:tau} dominates the relaxation
time. In particular, $\tau_\mathrm{rel}$ exhibits a
sharp maximum at $\delta = 0$, whose magnitude scales as
$\tau_\mathrm{rel} \propto \epsilon^{-3/2}$. This corresponds to
correlated dissociation events: neither of the two domain types  is in excess
with respect to the other, and so rebinding to a new partner is
conditioned on finding another thermally activated unbound domain within $v_\mathrm{cage}$. The concentrations of such unbound domains are
$c_\mathrm{A} = c_\mathrm{B} = \sqrt{K_d c_{\mathrm{AB}}}\propto \epsilon^{1/2}$. In
contrast, for $\delta \gg 1/v_\mathrm{cage}$ such that $n\gg 1$,
binding to a new partner is fast and essentially independent of
$\delta$, so that $\tau_\mathrm{rel} \propto \epsilon^{-1}$. This scaling
behavior is our central prediction, and is illustrated in
\Fig{fig:schematic}g.

\subsection*{Molecular dynamics simulations.}

\begin{figure*}[!hbt]
  \centering
  \includegraphics[width=\textwidth]{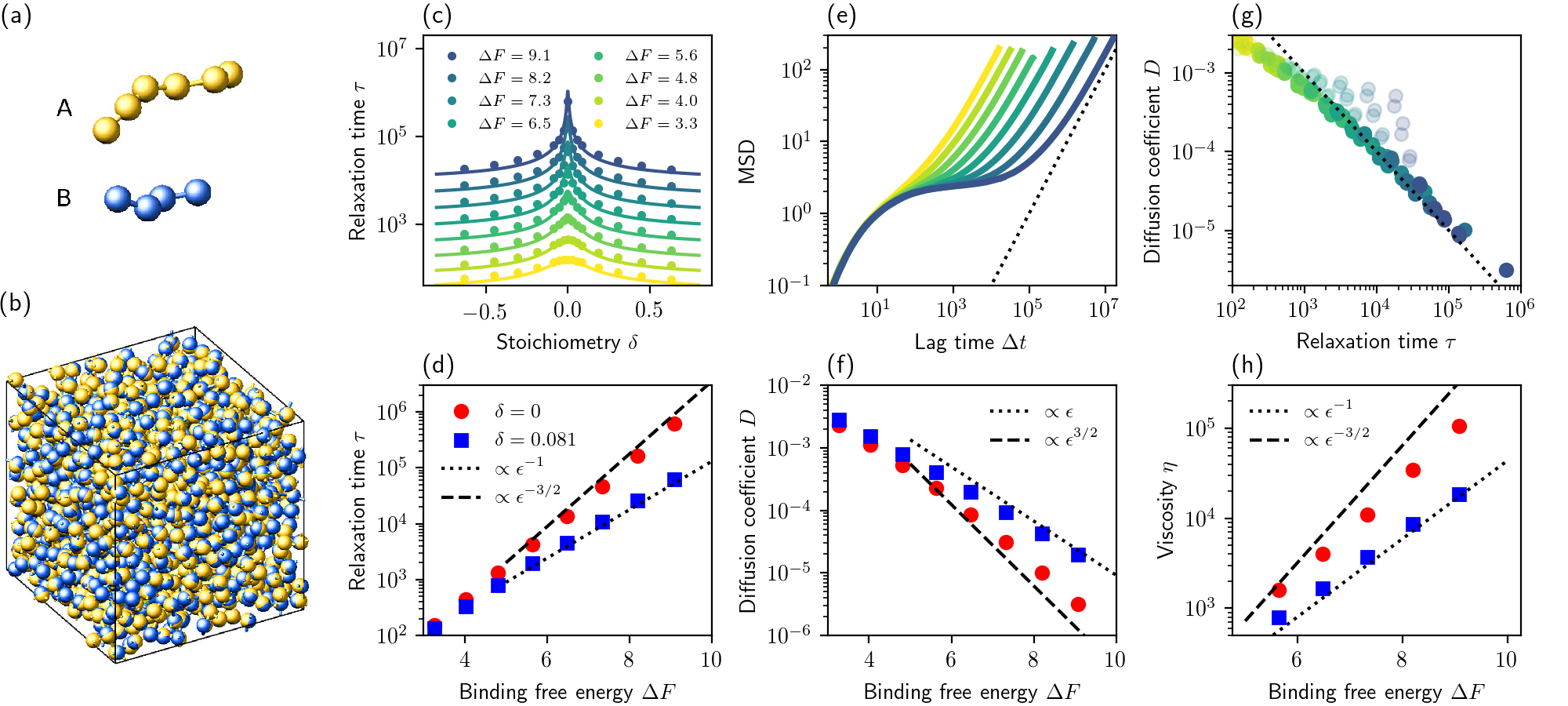}
  \caption{ \textbf{Molecular Dynamics simulations reveal the
      importance of stoichiometry to the dynamical properties of the
      condensate.} \textbf{(a)} MD model for the multivalent
    associative proteins. Colored spheres represent A and B
    domains. \textbf{(b)} Representative snapshot of the dense,
    network-forming liquid condensate. \textbf{(c)} Bond relaxation
    time (see text) as a function of stoichiometry for different
    binding strengths. Symbols indicate MD simulations; solid curves
    indicate theory (\Eq{eq:tau}) with $v_\mathrm{cage}=2.2$ (fitted,
    consistent with the plateau of MSD in (e)), $\tau_0 = 1.0$
    (corresponding to the unbinding time in the absence of any
    interaction), and
    $K_d= c_\mathrm{A} c_\mathrm{B} / c_{\mathrm{AB}}$ measured from
    data at $\delta = 0$. \textbf{(d)} Bond relaxation time
    $\tau_\mathrm{rel}$ as a function of binding strength is
    consistent with predicted scaling for both equal and unequal
    stoichiometries (\Eq{eq:tau}, \Fig{fig:schematic}g). \textbf{(e)}
    Mean squared displacement (MSD) of individual domains as a
    function of time reveals diffusive scaling (dashed line) at long
    times (here $\delta = 0$). \textbf{(f)} Diffusion coefficient of
    the minority species as a function of binding strength at equal
    and unequal stoichiometry. \textbf{(g)} Diffusion coefficient
    plotted against bond relaxation time, for all values of $\delta$
    and $\Delta F$. The dotted black line indicates
    $D\propto \tau_\mathrm{rel}^{-1}$. Transparent circles correspond
    to systems where one component is in large excess,
    $|\delta|>0.2 c_\mathrm{tot}$, for which disconnected proteins
    dominates the diffusivity.  \textbf{(h)} Viscosity, obtained using
    the Green-Kubo relation, as a function of binding strength, shows
    similar scaling to the bond relaxation time (d).  }
  \label{fig:MD}
\end{figure*}

We employ molecular dynamics simulations to test our theoretical
predictions for the relaxation time (\Eq{eq:tau}). Specifically, we
model the system schematized in \Fig{fig:schematic}a-b using a
bead-spring representation, where only the binding domains are
simulated explicitly (\Fig{fig:MD}a). Binding between complementary
domains is modeled by a soft attractive potential minimized when the
beads fully overlap, while strong repulsion between beads of the same
type prevents the formation of multiple bonds involving the same
domain (see Methods). The range of the repulsive interaction between
domains sets the unit of length, while the unit of time is chosen to be the
average time it takes for a free domain to diffuse a unit length. We
simulate only the dense phase of this phase-separating system
(\Fig{fig:MD}b). The control parameters are the binding free energy
$\Delta F$ and the stoichiometric difference $\delta = c_\mathrm{A} - c_\mathrm{B}$,
while the total concentration of domains $c_\mathrm{tot}$ is held
fixed. Simulations are performed using LAMMPS
~\cite{noauthor_lammps_nodate,plimpton_fast_1995} (see Methods).

We first study the relaxation of individual bonds. To quantify this
relaxation, we compute the bond adjacency matrix $A_{ij}(t)$, where
$A_{ij}(t) = 1$ if domains $i$ and $j$ are bound at time $t$, and $0$
otherwise. We first obtain the average autocorrelation function of
this matrix, $C(\Delta t) = \av{ \sum_{i,j} A_{ij}(t)A_{ij}(t+\Delta t)}_{t}$,
where the sum runs over all pairs of complementary domains, and then
extract the bond relaxation time $\tau$ by integration of the normalized autocorrelation,
$\tau = \int_0^\infty C(\Delta t) d\Delta t / C(0)$. The resulting relaxation time $\tau$ is
plotted as a function of stoichiometry difference
$\delta=c_\mathrm{A}-c_\mathrm{B}$ for different values of $\Delta F$
in \Fig{fig:MD}c (symbols). These values are in good agreement with
the theoretical prediction of \Eq{eq:tau} (solid curves), and in
particular exhibit a clear maximum at equal stoichiometry
($\delta = 0$). The magnitude and sharpness of the peak increases with
the binding free energy $\Delta F$. Furthermore, we confirm in
\Fig{fig:MD}d that $\tau$ scales as
$\epsilon^{-3/2} = \exp(3 \Delta F/2)$ at equal stoichiometry, and as
$\epsilon^{-1}=\exp(\Delta F)$ at unequal stoichiometry. Thus, the
relaxation time increases much faster with $\Delta F$ at equal
stoichiometry, in agreement with our
analytical prediction (\Eq{eq:tau}).

\subsection*{Diffusivity and viscosity.}

How does this sizable difference in relaxation times influence
macroscopic condensed-phase properties such as diffusivity and
viscosity? To answer these questions, we first monitor the mean
squared displacement (MSD) of individual binding domains as a function
of lag time (\Fig{fig:MD}e).  Several distinct regimes are apparent in
the MSD: short times correspond to bond-level vibrations, the plateau at intermediate
times reveals caging within the bonded network, while the long-time scaling
$\mathrm{MSD}\propto \Delta t$ is diffusive, confirming that the
system behaves as a liquid. We extract the long-time diffusion
coefficient from these simulations, and find that it directly reflects
the bond relaxation time, \emph{i.e.} $D\propto 1/\tau$
(\Fig{fig:MD}g), and thus scales as $\epsilon^{3/2}$ at equal
stoichiometry (\Fig{fig:MD}f). This shows that the slow bond relaxation within
the connected network dominates the diffusive properties of the
system. We note however that at large stoichiometry differences ($|\delta| > 0.2 c_\mathrm{tot}$, transparent symbols in
\Fig{fig:MD}g), fully
unbound proteins of the majority species exist and diffuse rapidly
through the network, thus violating these scaling laws.

Turning to the viscosity $\eta$ of the liquid, which we measure using
the Green-Kubo relation between viscosity and equilibrium stress
fluctuations~\cite{todd_nonequilibrium_2017}, we observe similarly
that $\eta \propto \tau$ (\Fig{fig:MD}h). The macroscopic transport
properties of this binary liquid thus directly reflects the highly
stoichiometry-dependent molecular relaxation mechanism illustrated in
\Fig{fig:schematic}: in the strong-binding regime, the viscosity of the
liquid dramatically increases near equal stoichiometry.

\subsection*{Mixing dynamics.}

\begin{figure}[bt]
  \centering
  \includegraphics[width=0.99\columnwidth]{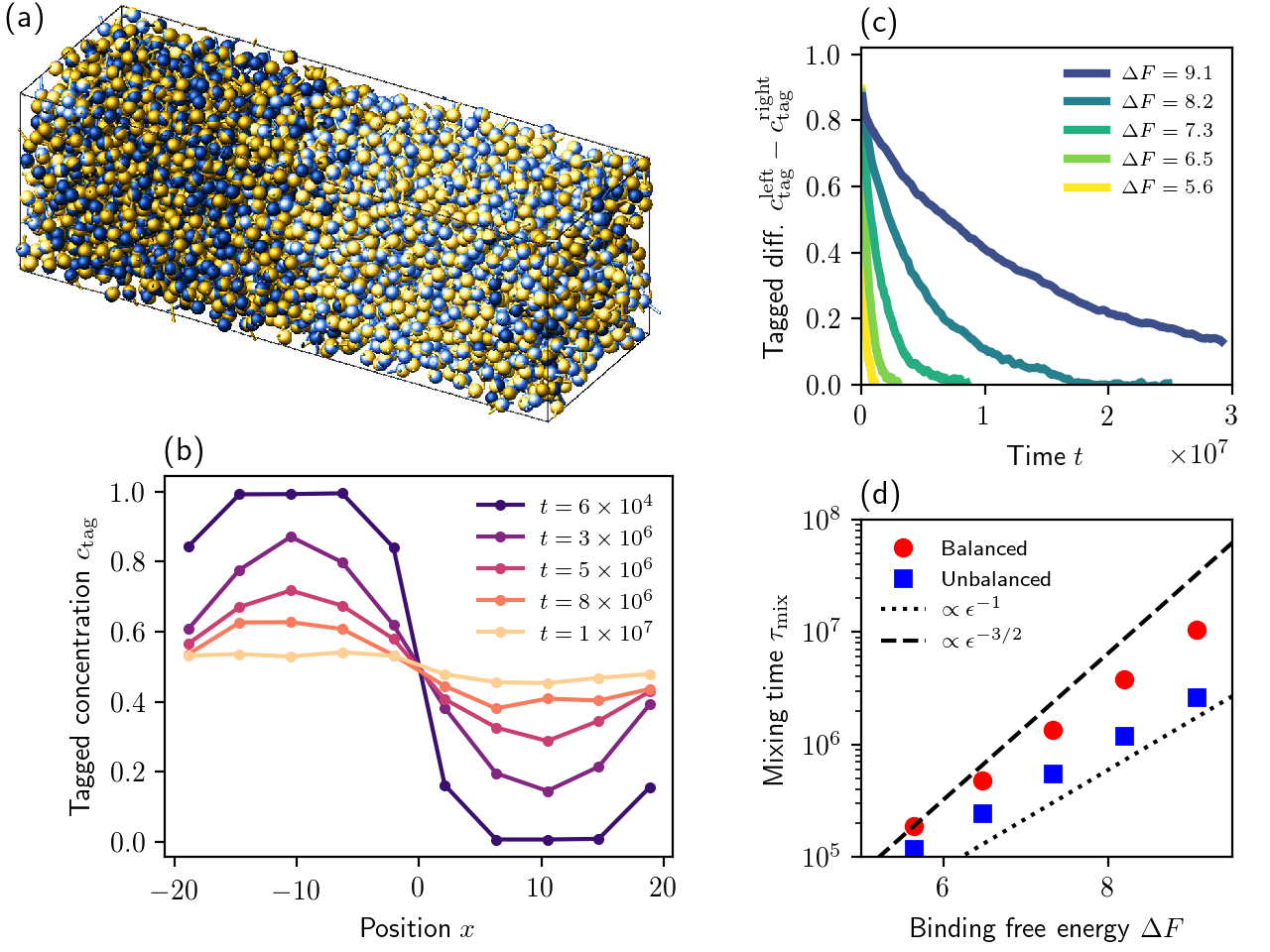}
  \caption{ \textbf{Composition controls mixing rate near equal
      stoichiometry.}  \textbf{(a)} Snapshot of an MD simulation with
    initially tagged particles on the left side of the
    box. \textbf{(b)} Concentration profiles for tagged particles
    along the long axis at different times, for equal stoichiometry
    $\delta = 0$, showing slow relaxation towards the homogeneous
    state. \textbf{(c)} Relaxation of the tagged concentration difference
    between the two half-boxes, for variable binding free
    energy. \textbf{(d)} Equilibration time as a function of binding
    strength. The unbalanced case has  $\delta = 0.061$. }
  \label{fig:mixing}
\end{figure}

Our predictions for the dependence of bulk transport coefficients on
the stoichiometry of the associative protein condensate have
experimentally testable consequences. For instance, by preparing an
homogeneous droplet and fluorescently tagging domain on one side, one
could measure the mixing dynamics as a function of the composition. We
simulate the relaxation of the composition profile for this case by
putting in contact two simulation boxes (\Fig{fig:mixing}a-b). We
monitor the relaxation of the tagged composition difference between
the two halves of the simulation box (\Fig{fig:mixing}c) and extract
the relaxation time by exponential fitting of the decay curve
(\Fig{fig:mixing}d). Consistent with our equilibrium analysis, we find
that mixing is much faster when a species is in excess
(\Fig{fig:mixing}d, squares) than when stoichiometry is balanced
(\Fig{fig:mixing}d, circles).  Interestingly, if the two boxes had
initially distinct compositions, mixing is significantly faster:
indeed, the gradient of bound fraction of the domain results in a
strong chemical potential gradient, and thus in a large thermodynamic
force restoring compositional homogeneity.

\subsection*{Discussion.}

In this Letter, we investigated the dynamics of protein-rich
condensates characterized by strong, specific interactions between
complementary binding sites. Our theoretical analysis of the
molecular-level relaxation mechanisms in these liquids suggests a
strong composition dependence: near equal stoichiometry of
complementary binding sites, the dynamics of the liquid dramatically
slows down. This slowing is due to the lack of free binding sites at
equal composition, which leads to a predominance of rebinding
following bond breaks. We confirmed this mechanism through molecular
dynamics simulations and showed that it controls the equilibrium
diffusivity and viscosity of the liquid network.

The molecular-level connectivity relaxation of protein liquids through
binding-unbinding events is generally not directly accessible in
experiments. By contrast, our predictions for macroscopic transport
quantities are readily testable, for instance using engineered protein
condensates such as SUMO-SIM~\cite{banani_compositional_2016} and
SH3-PRM~\cite{li_phase_2012} systems. Our predictions would also hold
in other liquids characterized by strong specific interactions, such
as in highly controllable DNA
nanoparticles~\cite{conrad_increasing_2019}. In such systems, the
effect of composition on diffusivity could be observed using
fluorescence recovery after
photobleaching~\cite{taylor_quantifying_2019} as in \Fig{fig:mixing} and nanoparticle
tracking~\cite{feric_coexisting_2016}, while our predictions on
viscosity and mixing dynamics could be tested by monitoring the shape
relaxation of merging droplets~\cite{ghosh_determinants_2020}.

While the dynamics of protein condensates can be regulated by many
factors, such as density~\cite{kaur_molecular_2019,ghosh_determinants_2020}, salt concentration, and the presence of RNA~\cite{elbaum-garfinkle_disordered_2015},
our work highlights the possibility that cells can also fine-tune the
mechanical and dynamical properties of their membraneless organelles
through small changes in composition. Beyond controlling the time
scale of internal mixing and merging of these droplets,
stoichiometry-dependent slowing could also impact the mobility
exchange rates of ``clients'' -- constituents of the condensates that
do not contribute directly to phase separation, but may be
functionally important for the cell~\cite{banani_biomolecular_2017}. Overall, we have shown that high
specificity liquids have unusual physical properties and provide novel
avenues that cells could use to regulate their phase-separated bodies.

\paragraph*{\emph{\textbf{Methods}}.}

Molecular Dynamics simulations are performed using the March 2020
version of LAMMPS~\cite{noauthor_lammps_nodate}. Proteins of type A
and B are represented by bead-spring multimers with respectively 6 and
4 binding domains (chosen with different valency to avoid magic-number
effects associated with the formation of stable
dimers~\cite{freeman_rosenzweig_eukaryotic_2017,xu_rigidity_2020,zhang_decoding_2020}). Simulations
are done in the NVE ensemble using a Langevin thermostat, with energy
normalized so that $k_B T = 1$.  Links between domains in a given
protein are modeled as finite extensible nonlinear elastic bonds, with
interaction potential
$E(r) = -0.5 K R_0^2 \log\left[1-(r/R_0)^2\right]$ as a function of
bond elongation $r$, with coefficients $K=3$ and $R_0=3$. Interaction
between domains of the same type are given by a repulsive truncated
Lennard-Jones potential,
$E(r) = 4\varepsilon
\left[\left(\frac{\sigma}{r}\right)^{12}-\left(\frac{\sigma}{r}\right)^{6}\right]$
with $\varepsilon = 1$, $\sigma = 1$ (which sets the unit of length),
and cutoff at $R=2^{1/6}$. Binding between complementary domains
occurs via a soft potential,
$E(r) = A \left(1+\cos(\pi r/r_c) \right)$ for $r<r_c$, with cutoff
$r_c = 0.5$. Energy is minimized when domains fully overlap, and
Lennard-Jones repulsive interaction between domains of the same type
ensured that binding is one-to-one. The interaction strength $A$ is
related to the binding free energy by
$\Delta F = - \ln \left(\int_{0}^{r_c} 4 \pi r^2 e^{-E(r)} dr / (4 \pi
  r_c^3/3)\right)$.  We set the average time it takes for an unbound
domain to diffuse a unit length to be the unit of time, $\tau_0 =
1$. The simulation step is $\delta t = 0.0176$.  We simulate only the
dense phase, with periodic boundary conditions (box size: $14^3$ for
\Fig{fig:MD}, $42\times 14\times 14$ for \Fig{fig:mixing}) and density
typical of a demixed droplet with free surface. The total
concentration $c_\mathrm{tot}=1.05$ of domains is kept fixed while the
stoichiometry $\delta$ is varied.

To ensure equilibration of the system, the attraction strength $A$ is
annealed from zero to its final value over a time of $5\tau$, where
$\tau$ is the bond relaxation time. The system then evolves for
another $5\tau$, prior to measurements performed over $20\tau$. In
\Fig{fig:MD}, measurements of $\tau$, MSD, and $D$ have $N=5$ repeats;
measurements of $\eta$ have $N=20$. Statistical error bars are smaller than
the symbol size. In \Fig{fig:mixing}, the system is initially annealed
with walls separating the two halves of the system, with different
labels for domains in either side.  At $t=0$, the walls
are removed and mixing starts.

\paragraph*{\emph{\textbf{Acknowledgments}}.}

This work was supported in part by the National Science Foundation,
through the Center for the Physics of Biological Function
(PHY-1734030).  

\bibliography{Stoichiometry}

\end{document}